\documentclass[aip,%
reprint, 
 amsmath,amssymb,
]{revtex4-2}

\usepackage{graphicx}
\usepackage{dcolumn}
\usepackage{bm}
\usepackage{xcolor}

\usepackage[utf8]{inputenc}
\usepackage[T1]{fontenc}
\usepackage{mathptmx}
\usepackage{etoolbox}

\usepackage{ulem}
\usepackage{soul}
\definecolor{matlabblue}{rgb}{0,0.447,0.741}
\definecolor{matlabred}{rgb}{0.85,0.325,0.098}

\makeatletter
\def\@email#1#2{%
 \endgroup
 \patchcmd{\titleblock@produce}
  {\frontmatter@RRAPformat}
  {\frontmatter@RRAPformat{\produce@RRAP{*#1\href{mailto:#2}{#2}}}\frontmatter@RRAPformat}
  {}{}
}%
\makeatother

\begin{document}

\preprint{APS/123-QED}

\title{A dissipative Nonlinear Schr\"{o}dinger model for wave propagation in {the marginal ice zone}}

\author{A. Alberello}
 \email{A.Alberello@uea.ac.uk}
\author{E. I. P\u{a}r\u{a}u}%
\affiliation{%
 School of Mathematics, University of East Anglia, Norwich, NR4 7TJ, United Kingdom}

\date{\today} 

\begin{abstract}
Sea ice attenuates waves propagating from the open ocean.
Here we model the evolution of energetic unidirectional random waves in {the marginal ice zone} with a nonlinear Schr\"{o}dinger equation, with a frequency dependent dissipative term consistent with current model paradigms and recent field observations.
The preferential dissipation of high frequency components results in a concurrent downshift of the spectral peak that leads to a less than exponential energy decay, but at a lower rate compared to a corresponding linear model.
Attenuation and downshift contrast nonlinearity, and nonlinear wave statistics at the edge tend to Gaussianity farther into {the marginal ice zone}.

\end{abstract}

\maketitle



Waves formed in the open ocean penetrate sea ice defining an extremely dynamical sea ice region, known as marginal ice zone (MIZ) \cite{wadhams1986}.
Around Antarctica, where large waves are generated all year round in the Southern Ocean, the MIZ can reach {hundreds} of kilometres \cite{stopa2018pnas,brouwer2021tcd} (with winter averages  $>200$\,km), therefore playing a substantial role in the climate system by regulating heat and momentum exchanges between ocean and atmosphere at large spatial scales \cite{massom2010ps,vichi2019grl}.
The urgent need to understand the often baffling sea ice trends \cite{turner2017nature,eayrs2021netgeo} prompted development of next-generation coupled waves and sea ice models and stimulated a surge of observational campaigns in recent years, e.g.\cite{thomson2018jgr,kohout2020ag}.
In {the MIZ}, the interplay between waves and sea ice activates feedback mechanisms that affect their respective properties.
On one hand, waves break-up large floes {\cite{kohout2014nature}} and limit the growth of the newly formed ones {\cite{roach2018jgr}}, also maintaining an unconsolidated sea ice cover \cite{womack2022ag}.
On the other hand, sea ice attenuates incoming waves via scattering and dissipation \cite{squire2020ar}, at different rates, depending on ice concentration, thickness and floes size.

Field measurements in the Arctic and Antarctic show exponential wave attenuation \cite{kohout2014nature,kohout2020ag}, and at a rate function of the wave period \cite{meylan2014grl,cheng2017jgr} (shorter waves attenuate faster).
For floes much smaller than the wavelength, as commonly found in the outer MIZ {\cite{alberello2019tc}}, the dominant attenuation mechanisms is assumed to be viscous wave dissipation \cite{keller1998jgr} (scattering is dominant when floes are of the same size of the wavelength\cite{squire2020ar}).
The ice cover is modelled as an homogeneous layer of constant, averaged, properties \cite{meylan2018jgr}, i.e.~individual floes are not resolved{, and viscosity is an effective viscosity (higher than molecular viscosity) that accounts for floes collisions, eddies, small scale turbulence and sea ice properties \cite{rabault2019jfm}}.
In this family of models, a wave dispersion relation with complex wave number is found, in which the imaginary part defines the frequency dependent wave attenuation rate.
A power law attenuation with respect to the wave frequency is recovered \cite{meylan2018jgr}, with scaling depending on the physical dissipation mechanism.

The Nonlinear Schr\"{o}dinger Equation (NLS), an universal model for weakly nonlinear waves dynamics in the open ocean
{\cite{onorato2013pr}} (despite not reproducing wave breaking), is also widely used to describe wave propagation in the presence of dissipation {\cite{segur2005jfm,dias2008pla,onorato2012pla}}.
The dissipative NLS (dNLS) includes a damping term, due to viscosity or other sources, that matches the decay rate of the linear wave amplitude \cite{dias2008pla,wu2006jfm}.
Damping stabilises the modulational instability, to which the NLS is naturally subjected, by diminishing the unstable region of disturbances \cite{wu2006jfm}.
It is worth noting that a NLS-type equation for waves in sea ice has been derived by \citet{liu1988jpo}, but for waves propagating on an ocean covered by an elastic plate, and better represent compact ice conditions (floes much larger than the waves) towards the interior of the MIZ.

Here we propose a dNLS for waves in {the MIZ}, in which the damping term is derived from current sea ice model paradigms for {a sea ice zone} comprised of small floes (relative to wavelength) and for which viscous losses are the main dissipation mechanism \cite{squire2020ar}.
We study the dynamics of 
energetic waves, using {random} simulations for two sea ice conditions (corresponding to low and high dissipation respectively as defined by the recent empirical formulations derived by \citet{kohout2020ag}) that qualitatively reproduce field observations during Southern Ocean storm conditions \cite{kohout2014nature,kohout2020ag,alberello2021arxiv}.
We show that the apparent downshift of the energy, due to the higher attenuation of shorter waves, slows down the total wave energy decay that, as a consequence, deviates from the exponential behaviour.

The NLS for surface gravity waves propagating in space, i.e. analogous to the hydrodynamic wavemaker problem \cite{chabchoub2016fluids}, in deep water conditions and without dissipation is:
\begin{equation}
\label{eqn:nls}
i\left(\dfrac{\partial B}{\partial x}+\dfrac{1}{c_{g}}\dfrac{\partial B}{\partial t}\right)-\dfrac{1}{g}\dfrac{\partial^2B}{\partial t^2}-k^3|B|^2B=0,
\end{equation}
where $B(x,t)$ is the slowly varying complex amplitude of the surface elevation in space ($x$) and time ($t$), $c_{g}$ and $k$ the group speed and wavenumber of the carrier wave (the fast oscillation), and $g$ is the gravitational acceleration.
The boundary condition is defined by $B_0=B(x_0,t)$.

Attenuation is introduced in Eqn.~\ref{eqn:nls} as an additional damping term using an heuristic approach \cite{dias2008pla}.
In open water, damping formally derives from adding viscous dissipation to the dynamic free surface boundary condition of the standard potential flow problem \cite{wu2006jfm}.
Viscous models of the form $\phi$ and $\partial^2\phi/\partial z^2$ ($\phi$ denotes the free surface velocity potential and $z$ the vertical coordinate) have been proposed \cite{wu2006jfm}.
In sea ice, damping has the same form of viscous dissipation but derives from the additional ice-induced pressure at the free surface \cite{meylan2018jgr}.
In a recent review, \citet{meylan2018jgr} proposed a model paradigm with ice-induced pressure of the form $\partial\phi/\partial z$ (derived assuming that the wave energy loss is proportional to the square horizontal velocity times the ice thickness $h_i$) that leads to a dispersion relation in which the imaginary part of the wave number, i.e.~the one defining the exponential attenuation rate, {is}:
\begin{equation}
\label{eqn:ki}
k_I = \dfrac{h_i\rho_i\nu}{\rho_w g^2}\omega^3,
\end{equation}
where $\omega$ is the wave angular frequency, $\rho_i$ and $\rho_w$ the ice and water density, and $\nu$ a property of the sea ice (in $s^{-1}$).
The real part of the dispersion relation coincides with the open water one ($k=\omega^2/g$), therefore group velocity does not change in response to dissipation, in agreement with measurements in thin ice for wave components longer than $\approx5$\,s \cite{cheng2017jgr}.
Other model paradigms exist that lead to different dependency on the angular frequency (of the type $\omega^n$), but this model is chosen because the $\omega^3$ dependency agrees with field measurements \cite{meylan2018jgr}.
Adding dissipation to Eqn.~\ref{eqn:nls}
yields to the dNLS:
\begin{equation}
\label{eqn:dnls}
i\left(\dfrac{\partial B}{\partial x}+\dfrac{1}{c_{g}}\dfrac{\partial B}{\partial t}\right)-\dfrac{1}{g}\dfrac{\partial^2B}{\partial t^2}-k^3|B|^2B=-i\dfrac{h_i\rho_i\nu}{\rho_wg^2}\omega^3B.
\end{equation}

To reproduce Southern Ocean storm waves,
without loss of generality,
the initial energy distribution is defined by a Gaussian swell spectrum \cite{lucas2015oeng}:
\begin{equation}
\label{eq:spec}
|\hat{B}_0(\omega)|^2 = \dfrac{a^2}{\sigma \sqrt{2\pi}}\exp{\left\{-\dfrac{1}{2}\left(\dfrac{\omega}{\sigma}\right)^2\right\}}
\end{equation}
where $\sigma$ denotes the width of the spectrum and $a^2$ a scaling parameter proportional to the wave energy.
The initial, linear, condition is obtained from amplitudes directly extracted from the spectrum and phases randomly generated with a uniform distribution in $[0,2\pi)$.
To match waves at the ice edge reported in \citet{alberello2021arxiv},
$a$ is chosen to give significant wave height $H_S=7.3$\,m ($H_S=4\sqrt{m_0}$, $m_0$ is the zeroth order moment of the spectrum) and
the peak wave period is set to $T_0=12$\,s ($\omega_0=2\pi/T$; the corresponding wavelength is $\lambda=225$\,m).The characteristic wave steepness ($\varepsilon=\pi H_S/\lambda$), is 0.10, a typical value for storm waves in the Southern Ocean.
Spectral width is set to $\sigma=\omega_0/8$.

Two sea ice conditions are tested, corresponding to low and high dissipation respectively, to qualitatively mimic wave height decay
for low ($<80\%$) and high ($>80\%$) sea ice concentration in the MIZ reported by \citet{kohout2020ag}, i.e.~wave height at 50\,km (or $\approx200$ wavelengths) is $\approx75\%$ and $\approx20\%$ of the open ocean one for waves shorter than 14\,s.
The empirical formulation by \citet{kohout2020ag}, based on a large dataset of recent field observations in the Southern Ocean MIZ, provides a suitable benchmark for our simulations.
The two dissipation regimes are achieved by maintaining constant and homogeneous ice thickness ($h_i=0.3$\,m), ice and water density ($\rho_i=900$\,kg\,m$^{-3}$ and $\rho_w=1027$\,kg\,m$^{-3}$), and varying $\nu$ from 0.02 to 0.2\,s$^{-1}$ (low to high dissipation).
Ice concentration is implicitly incorporated by $\nu$, which can be rewritten as the product of $\nu_{100}$ (in 100\% sea ice) and ice concentration, i.e.~$\nu = \nu_{100}c_i$.
Water and sea ice properties are typical of the MIZ \cite{alberello2020jgr}.
The parameter $\nu$ relates to the viscous losses but its physical meaning is more uncertain and not directly measured in the field \cite{mosig2015jgr}, and usually inferred from model inversion \cite{rogers2016jgr}.
It is worth noting that the exterior of the MIZ, which we model, is formed by floes much smaller than the characteristic wavelength of Southern Ocean waves (tens of meters versus hundreds of meters), and therefore viscous losses dominate \cite{squire2020ar}. Reflection at the sea ice edge are instead linked to scattering, when wavelength and floe size are of the same order \cite{squire2020ar}, which is not the case for conditions considered in this Letter.

Eqn.~\ref{eqn:dnls} is solved advancing $B(x,t)$ in space using a fourth order Runge-Kutta method. The time derivatives are efficiently computed in the Fourier space.
Careful considerations should be given to the solution of $\omega^3B$ (dissipative term). The shift between the spectrum of the slowly varying envelope (centred around $\omega=0$) and the fast oscillation (centred around $\omega=\omega_0$) should be reintroduced in the Fourier space.
Coefficients $c_{g}$ and $k$ are updated at each spatial step to account for spectral changes.
Simulations are performed over a 55\,km domain with resolution $\Delta x=1$\,m.
The first 5\,km reproduce open ocean to allow the development of wave nonlinearity from the initially linear sea state over a distance of $\approx20$ wavelengths, therefore creating more realistic waves at the sea ice edge.
The transition to sea ice is achieved by activating dissipation in sea ice only (equivalent to using Eqn.~\ref{eqn:nls} in open ocean).
A periodic temporal domain of $512T_0$ ($t=6144$\,s) is used, discretised in $2^{12}$ elements ($\Delta t=1.5$\,s).
Temporal and spatial resolution guarantee numerical stability, and conservation of the wave energy in the open ocean.
To obtain statistically robust results, for each of the two sea ice conditions, 10 realisations are generated.
The amount of data generated allows to reliably assess probability levels as low as $2\times10^{-4}$, a larger number of simulations is only needed to investigate even lower probability levels.

\begin{figure}
  \centering
  \includegraphics[width=1.00\linewidth]{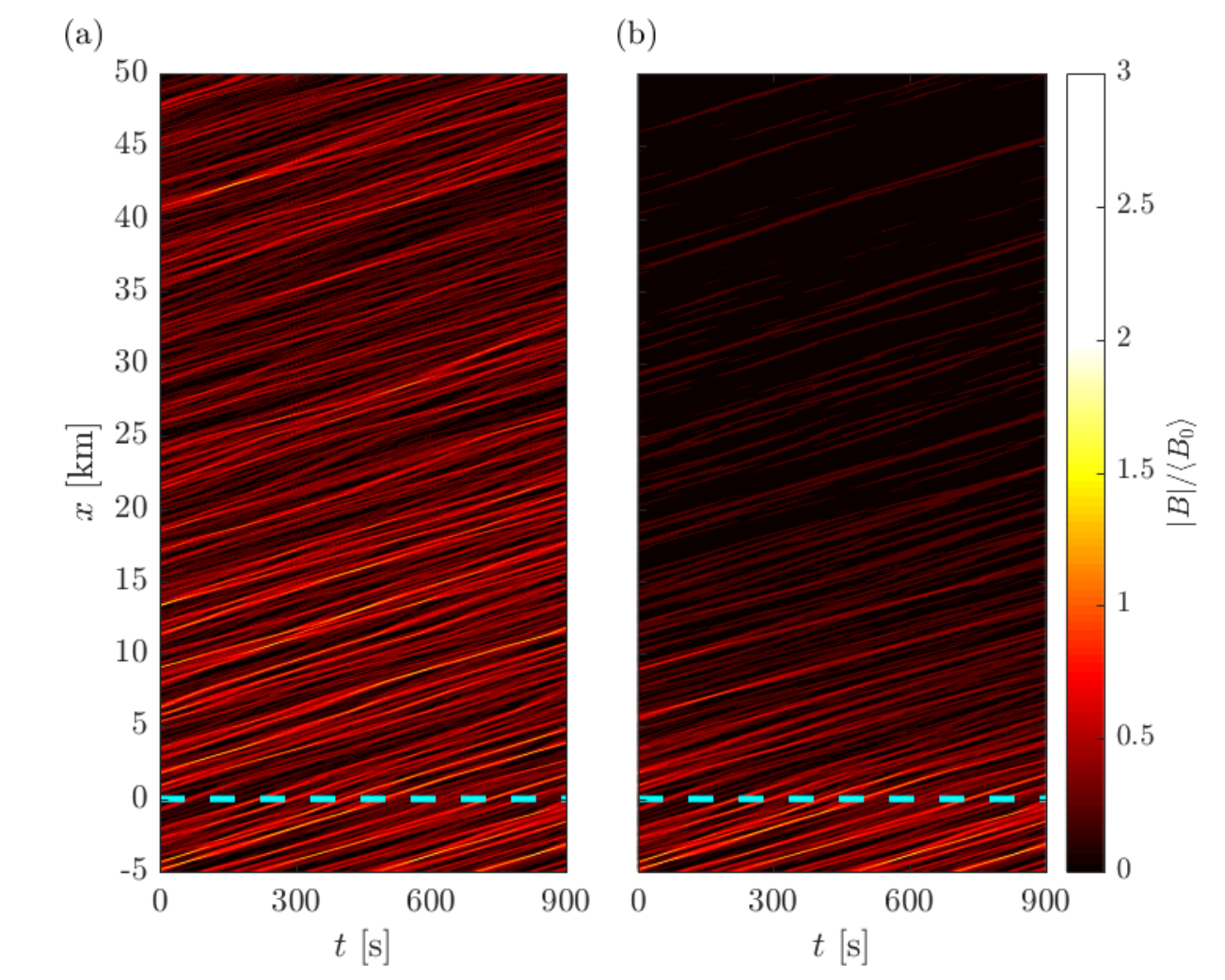}
  \caption{Space time evolution of the wave envelope for low (a) and high dissipation (b). The dashed line denotes the sea ice edge.}
  \label{fig:fig1}
\end{figure}

A sample of the wave surface elevation, expressed by the modulus of the wave envelope normalised by the mean wave amplitude ($|B|/\langle B_0\rangle$), is shown in Fig.~\ref{fig:fig1}a and b for low and high dissipation respectively.
In open ocean ($x<0$\,km) the evolution is governed by the NLS (Eqn.~\ref{eqn:nls}), and in sea ice ($x>0$\,km) by the dNLS (Eqn.~\ref{eqn:dnls}).
In the low dissipation regime energetic wave groups are detected deep in sea ice (amplification $|B|/\langle B_0\rangle>1$; Fig.~\ref{fig:fig1}a), albeit less frequently than at the edge.
The oblique lines that highlight energetic wave groups (the slope of which corresponds to the group speed) become sparser farther from the ice edge.
In the high dissipation regime wave groups decay faster (Fig.~\ref{fig:fig1}b), and $|B|/\langle B_0\rangle<1$ for $x>20$\,km.

\begin{figure}
  \centering
  \includegraphics[width=1.00\linewidth]{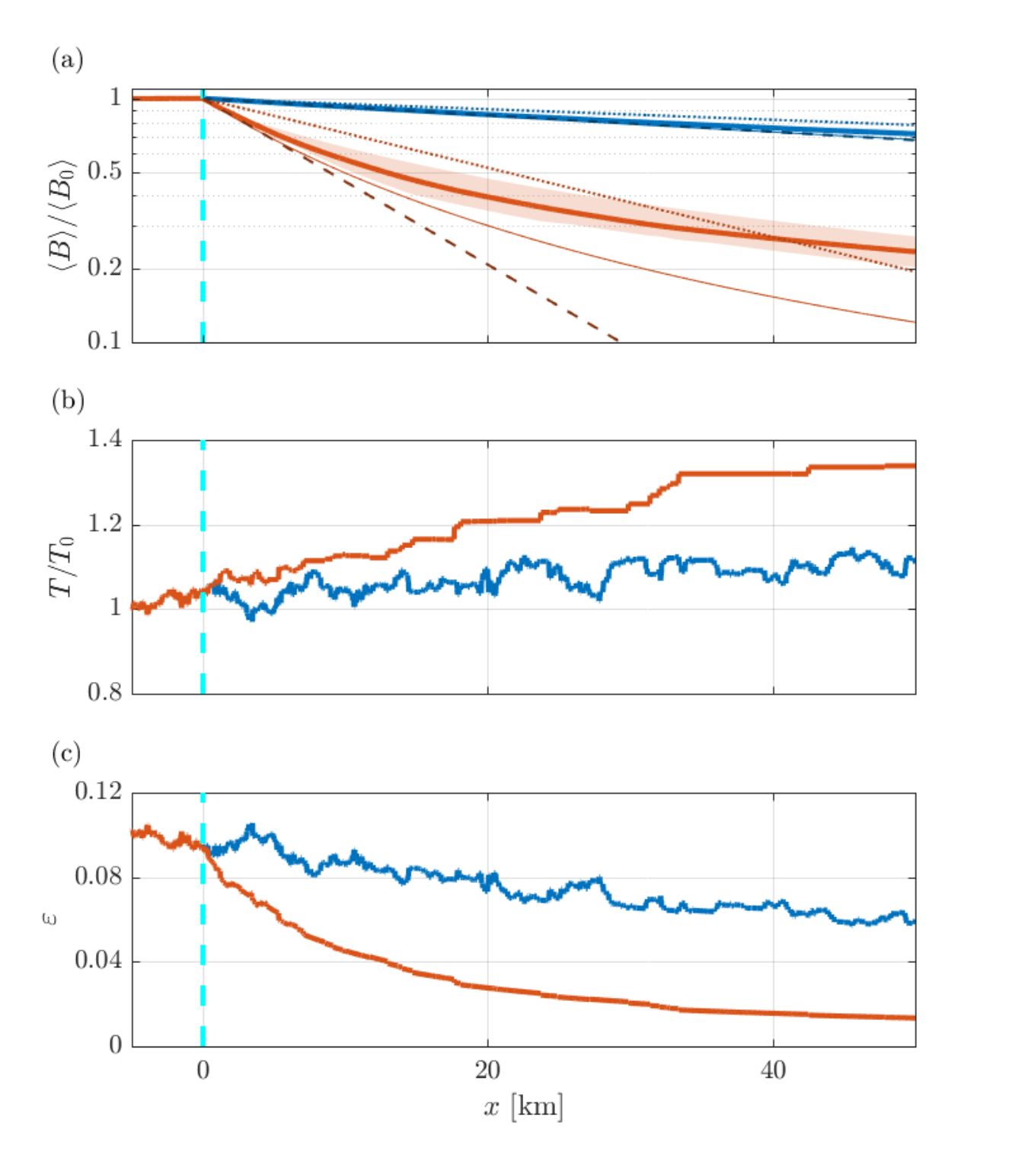}
  \caption{Spatial evolution of the wave amplitude (a; in logarithmic scale){,} peak period (b) and wave steepness (c) for low (blue) and high dissipation (orange). The average from the {random }simulations are denoted by a solid line (shaded area in (a) denote range from all simulations), and the dashed vertical line sea ice edge.
  In (a), the thinner line indicates linear predictions, and the dashed line the exponential decay at the initial peak ($k_{I,l}=7.8\times10^{-6}$\,m$^{-1}$ and $k_{I,h}=78.5\times10^{-6}$\,m$^{-1}$) and the dotted line from \citet{kohout2020ag} ($k_{K,l}=5.0\times10^{-6}$\,m$^{-1}$ and $k_{K,h}=32.7\times10^{-6}$\,m$^{-1}$).
   }
  \label{fig:fig2}
\end{figure}

The spatial evolution of the dimensionless wave amplitude is shown in Fig.~\ref{fig:fig2}a. 
The energy is conserved over the first 5\,km of propagation (in the open ocean), confirming the robustness of the numerical code.
For low dissipation (blue line), the wave amplitude decays to 0.716 at 50\,km almost exponentially, i.e.~a straight line with the $y$-axes in logarithmic scale.
The scatter amongst the 10 simulations, denoted by the shaded area, remains narrow (from 0.705 to 0.743 at 50\,km).
For high dissipation (red line), the attenuation is more substantial and the trend less than exponential, denoting a reduction of the decay rate as wave progress in sea ice.
At 50\,km the wave amplitude is 0.235 and the scatter almost doubles compared to the low dissipation case (from 0.203 to 0.272).
It is not surprising that differences in wave attenuation between low and high dissipation are of the same order of magnitude of variation in $\nu$.

For reference, dNLS simulations are compared against solutions obtained from a linear model with the same frequency dependent attenuation rate and with the same conditions at $x=0$\,km, i.e.~$\hat{B}_{lin}(x;\omega)=\hat{B}(x=0;\omega)\exp{(-k_Ix)}$.
The linear model (thin line in Fig.~\ref{fig:fig2}a) also displays a less than exponential decay,
however the dNLS is less dissipative than the corresponding linear model, indicating that nonlinearity slows down the energy attenuation, likely by shifting energy to less dissipative wave components at lower frequencies.
For low dissipation the difference between linear and nonlinear model at $x=50$\,km is $\approx0.04$, but for high dissipation is more substantial and the dNLS predicts double the amplitude ($\approx0.24$ versus $\approx0.12$).

Attenuation obtained from the dNLS simulations is also benchmarked against predictions based on constant exponential rates at the initial peak (Eq.~\ref{eqn:ki} for $T=12$\,s gives $k_{I,l}=7.8\times10^{-6}$\,m$^{-1}$ and $k_{I,h}=78.5\times10^{-6}$\,m$^{-1}$ which correspond to dissipation lengthscale of $800$\,km and $80$\,km respectively; subscript $h$ and $l$ stand for high and low dissipation) and parametrisations based on field data \cite{kohout2020ag} ($k_{K,l}=5.0\times10^{-6}$\,m$^{-1}$ and $k_{K,h}=32.7\times10^{-6}$\,m$^{-1}$ for $T<14$\,s; dissipation lengthscale of 1250\,km and 190\,km respectively)
that do not capture the slow-down of attenuation.
Note that low and high dissipation in \citet{kohout2020ag} correspond to sea ice concentration lower and higher than 80\%, whereas the simulations implicitly account for it in $\nu$.
The attenuation rates at the peak (dashed lines in Fig.~\ref{fig:fig2}a) result in lower wave amplitudes compared to the dNLS simulations (and also the linear model).
For low dissipation, the difference between attenuation at the peak and simulations is small, 0.675 versus 0.716 at 50\,km, but
for high dissipation the difference is more conspicuous, and the residual wave amplitude at 50\,km differs by one order of magnitude (0.02 versus 0.235).
By design, the attenuation by \citet{kohout2020ag} (dotted lines in Fig.~\ref{fig:fig2}a) closely matches the dNLS simulations at 50\,km but under-predicts attenuation rates close to the sea ice edge and over-predicts them farther into the sea ice, particularly in high dissipation, because of its constant attenuation rate.

The wave peak period increases in sea ice (Fig.~\ref{fig:fig2}b).
For low dissipation peak period increases by $\approx10\%$ at 50\,km,
and for high dissipation by $\approx35\%$.
Most of the increase is attributed to the stronger attenuation of short period waves \cite{meylan2018jgr}.
It should be noted that already at the sea ice edge $T/T_0>1$ because the NLS reproduces nonlinear wave-wave interactions leading to the downshift of the spectral peak and the energy cascade towards the high frequency tail \cite{dysthe2003jfm}.

The attenuation rate at the sea ice edge from the dNLS matches the one at the initial peak ($k_{I,l}$ and $k_{I,h}$), i.e.~the curves have the same slope.
The increasing peak period in sea ice results in a lower attenuation of the dominant wave component ($k_I\propto\omega^3=1/T^3$), and
for strong dissipation the $\approx35\%$ increase corresponds to $\approx2.45$ times weaker attenuation rate at the peak {after} 50\,km (or $\approx$200 wavelengths).
To a certain extent, the shift towards higher peak periods explains the slow-down of the wave energy decay deeper in sea ice, suggesting that the dissipation at the peak can be representative of the entire spectral attenuation if the peak period evolution is also reproduced.

Wave dissipation and peak period increase both contribute to the reduction of the wave steepness ($\varepsilon$) in sea ice (Fig.~\ref{fig:fig2}c). In the less dissipative sea ice regime the steepness is reduced by $\approx40\%$ over the 50\,km propagation, from 0.10 to 0.06.  The reduction is more substantial in the more dissipative regime ($\approx85\%$; from 0.10 to 0.015). 
The breaking probability, already low for $\varepsilon=0.10$ \cite{toffoli2010grl}, is further reduced in sea ice (field observations under conditions similar to those of the simulations revealed no breaking in the MIZ \cite{alberello2021arxiv}), meaning that the weakly nonlinear model is a suitable tool to investigate wave dynamics in the MIZ.
Reduction of wave nonlinearity deeper in the MIZ, particularly in the high dissipation regime, also means that dissipation contributes to stabilising modulational instability.

\begin{figure}
  \centering
  \includegraphics[width=1.00\linewidth]{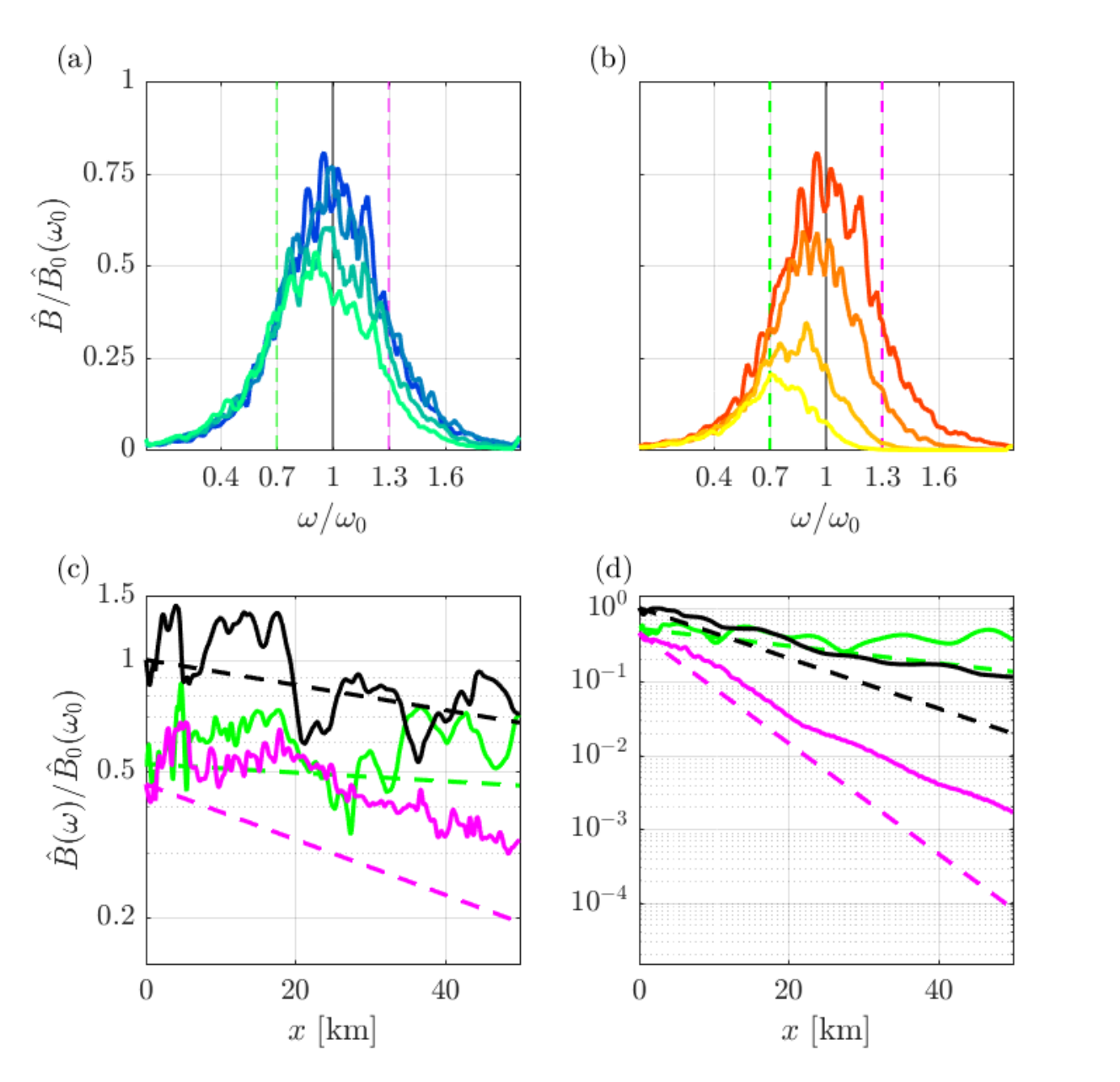}
  \caption{Wave spectra at progressive distances from the edge ($x=0,5,20,50$\,km; blue to cyan, and red to yellow) for low (a) and high dissipation (b). Amplitudes of the modes at $\omega/\omega_0=$ 0.7 {(green)}, 1.0 {(black)}, 1.3 {(magenta)} for low (c) and high dissipation (d). Dashed lines denote linear prediction (Eq.~\ref{eqn:ki}).}
  \label{fig:fig4}
\end{figure}

The wave spectral evolution in sea ice is shown in Fig.~\ref{fig:fig4}a--b, for low and high dissipation respectively.
Long waves (to the left of the peak) undergo low dissipation, even in the more dissipative case, in contrast to short waves (to the right of the peak).
Whereas most of the spectral changes can be attributed to the frequency dependent dissipation, consistent with field \cite{alberello2021arxiv} and laboratory measurements \cite{alberello2021ijope}, wave nonlinearity also contribute to energy exchanges between energy modes {\cite{annekov2006prl}}, albeit its effect weakens when wave energy content decays.

Evolution of individual modes is shown in Fig.~\ref{fig:fig4}c--d, for low and high dissipation respectively.
Despite the imposed attenuation is $\propto \omega^3$, the modes deviate from this scaling (the lines depart from {the} straight dashed lines
in {the} logarithmic plot 
that denote linear predictions from \citet{meylan2018jgr}){, particularly for high dissipation (Fig.~\ref{fig:fig4}d)}.
{The modes more closely follow the trend given by linear prediction for low dissipation (Fig.~\ref{fig:fig4}c).}
At low frequencies ({green line}), the amplitude oscillates around an almost constant value, hinting at the existence of nonlinear wave-wave energy transfers.
At high frequency ({magenta line}), the modes decay fast but the less than exponential trend suggests the presence of an energy input, likely due to the nonlinear energy cascade towards high frequencies \cite{fadaeiazar2018wm,fadaeiazar2020pre}, that contrasts dissipation.
A more technical tricoherence analysis \cite{campagne2019prf} would be needed to better quantify the energy cascade due to the four wave interactions in the presence of dissipation.
Here we note that, in the low dissipation case, the mode at $\omega/\omega_0=0.7$ (equivalent to $T=17\,s$) approximately equates the amplitude of the mode $\omega/\omega_0=1$ (equivalent to $T=12\,s$) after 35\,km (Fig.~\ref{fig:fig4}c), and likely exceeds it for $x>50$\,km, due to the combined effect of dissipation and nonlinearity. In the high dissipation case, the process is much faster, and the $\omega/\omega_0=0.7$ mode exceeds the $\omega/\omega_0=1$ for $x>15$\,km (Fig.~\ref{fig:fig4}d).

\begin{figure}
  \centering
  \includegraphics[width=1.00\linewidth]{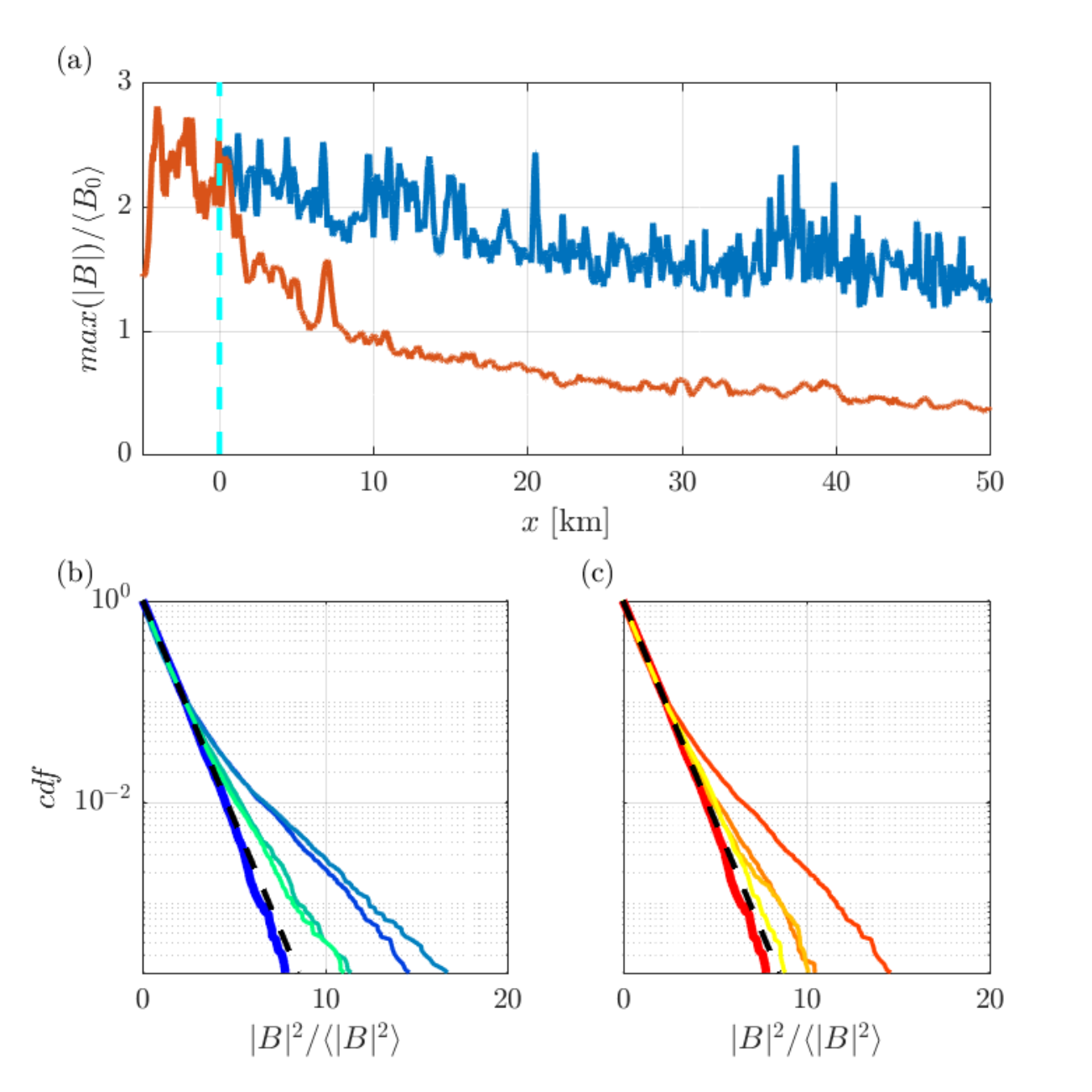}
  \caption{Maximum surface elevation in space (a) for low (blue) and high dissipation (orange). Exceedance probability at $x=-5,0,5,20,50$\,km for low (b) and high dissipation (c) truncated at $2\times10^{-4}$ for which the probability level are assessed reliably; the thicker line is for $x=-5$\,km, lighter shades are at progressive distances from the edge (blue to cyan, and red to yellow), and the black dashed line the exponential distribution.}
  \label{fig:fig3}
\end{figure}

In the ocean, wave nonlinearity leads to the formation of large individual waves \cite{onorato2013pr}.
The maximum surface elevation in the dNLS is tracked in space, see Fig.~\ref{fig:fig3}a. 
In the ice free portion of the domain the wave amplification approaches 3, as predicted for the Peregrine breather \cite{chabchoub2016prl,alberello2018oeng}.
In sea ice large waves decrease in amplitude, similarly to the trend observed for the wave amplitude (see Fig.~\ref{fig:fig2}).
For low dissipation, individual waves exceeding $2\langle B_0\rangle$ are detected up to $\approx40$\,km into sea ice.
For high dissipation, the largest waves drop below $\langle B_0\rangle$ after $\approx10$\,km of sea ice.

The occurence of exceptionally large waves at distance from the sea ice edge is analysed via the exceedance probability ($cdf$; Fig.~\ref{fig:fig3}b--c).
For linear sea states the wave amplitudes are Rayleigh distributed {\cite{janssen2014jfm}}, and $|B|^2$ is exponentially distributed (black dashed line in Fig.~\ref{fig:fig3}b--c).
At $x=-5$\,km (thick line in Fig.~\ref{fig:fig3}b--c) the initial condition is given as a linear superposition of modes, and its distribution closely resembles the benchmark exponential.
The wave propagation with no dissipation in open ocean (from $x=-5$\,km to $x=0$\,km) allows for nonlinear energy exchanges between modes that lead to the formation of rogue waves and, consequently, a departure of the $cdf$ from the exponential.
In sea ice (for $x\geq0$\,km), loss of wave energy and concurrent downshift of the spectral peak both contribute to the reduction of wave nonlinearity ({wave steepness; see Fig.\ref{fig:fig2}c}).
Waves become more linear and the $cdf$ tends to the exponential (lighter shades of blue/red in Fig.~\ref{fig:fig3}b--c).
For low dissipation, the exceedance probability maintain a deviation from the exponential even after 50\,km (cyan line Fig.~\ref{fig:fig3}b) due to the high wave energy deep in sea ice.
For high dissipation, the more substantial energy loss leads to an almost complete suppression of wave nonlinearity and the $cdf$ at $x=50$\,km (yellow line Fig.~\ref{fig:fig3}c) returns to the initial exponential distribution.

In summary, we proposed a model for wave propagation in {the MIZ} based on the NLS framework, by introducing a frequency dependent attenuation derived from viscous sea ice models that have been verified against field measurements \cite{meylan2018jgr}, and used to investigate the dynamics of energetic storm waves ($H_S=7.3$\,m; $T=12$\,s).
The problem setup reproduces typical Southern Ocean MIZ conditions \cite{kohout2020ag,alberello2021arxiv}, where waves are longer than the characteristic floe size and wave breaking is absent.
Stronger attenuation of high frequency components compared to low frequency ones leads to a downshift of the spectral peak (conspicuous in high dissipation regime) and a less than exponential wave attenuation, in contrast to predictions based on the total wave energy that do not account for the spectral downshift (cf.~\citet{kohout2020ag}).
Dissipation dominates over nonlinearity, but the latter contributes to shifting energy to more conservative modes via wave-wave interactions (noting that a more detailed tricoherence analysis\cite{campagne2019prf} is needed) and, as a result, partially counters attenuation further slowing down the energy decay, i.e.~waves in the dNLS are larger than those predicted by the linear model with same attenuation parameters.
In high dissipation regime, the residual wave amplitude 50\,km into sea ice ($\approx200$ wavelengths of propagation) for nonlinear waves is double the one of linear ones (Fig.~\ref{fig:fig2}a; $H_{S,nlin}= 1.75$\,m versus $H_{S,lin}= 0.88$\,m from $H_S=7.3$\,m at the sea ice edge).
The probability of large waves tends to Gaussianity farther into sea ice, due to concurrent attenuation of energy and downshift of the peak that reduce wave nonlinearity.
Unfortunately the model cannot be predictive, unless the viscosity parameter $\nu$ is linked to observable sea ice properties (the model can help revealing the nature of $\nu$ when paired with field observation), but these results prompt to the re-analysis of existing dataset to evaluate attenuation coefficients in sea ice and asses the role of wave nonlinearity.

Ultimately, the proposed dNLS indicates that the MIZ is wider than a linear model would predict, due to the higher residual wave energy that can break large floes and maintain the small ones unconsolidated farther from the sea ice edge.
{Heat and momentum exchanges in sea ice regions are altered, with consequences on the MIZ dynamics.}
These results can inform the development of the next-generation wave and sea ice coupled models as well as the planning of new observational campaign.

\begin{acknowledgments}
The research presented in this paper was carried out on the High Performance Computing Cluster supported by the Research and Specialist Computing Support service at the University of East Anglia.
\end{acknowledgments}

\section*{Author Declarations}
\subsection*{Conflict of Interest}
The authors report no conflict of interest.

\section*{Data Availability Statement}
The data that support the findings of this study are available from the corresponding author upon reasonable request.

\section*{references}


%

\end{document}